\documentclass[conference,a4paper]{IEEEtran}
\ifCLASSINFOpdf
  \usepackage[pdftex]{graphicx}
\else
  \usepackage[dvips]{graphicx}
\fi
\usepackage[table]{xcolor}
\usepackage{balance}
\usepackage{tabularx}
\usepackage{amsmath,amssymb}
\ifCLASSOPTIONcompsoc
 \usepackage[caption=false,font=normalsize,labelfont=sf,textfont=sf]{subfig}
\else
 \usepackage[caption=false,font=footnotesize]{subfig}
\fi
\newcommand{\rot}[0]{\scriptscriptstyle \mathrm{ROT}}
\newcommand{\ot}[0]{\scriptscriptstyle \mathrm{RO}}
\newcommand{\oo}[0]{\scriptscriptstyle \mathrm{OO}}
\newcommand{\us}[0]{\scriptscriptstyle \mathrm{US}}
\newcommand{\ro}[0]{\scriptscriptstyle \mathrm{RO}}
\newcommand{\sot}[0]{\scriptscriptstyle \mathrm{SOT}}
\newcommand{\Ss}[0]{\scriptscriptstyle \mathrm{SS}}
\newcommand{\sos}[0]{\scriptscriptstyle \mathrm{SOS}}
\newcommand{\ris}[0]{\scriptscriptstyle \mathrm{RIS}}
\newcommand{\rl}[0]{\scriptscriptstyle \mathrm{RL}}
\newcommand{\rs}[0]{\scriptscriptstyle \mathrm{RS}}
\newcommand{\ST}[0]{\scriptscriptstyle \mathrm{ST}}
\newcommand{\RT}[0]{\scriptscriptstyle \mathrm{RT}}
\newcommand{\tg}[0]{\scriptscriptstyle \mathrm{TG}}
\newcommand{\ros}[0]{\scriptscriptstyle \mathrm{ROS}}

\newcommand{\change}[1]{{\color{black}#1}}
\makeatletter
\let\oldabs\abs
\def\abs{\@ifstar{\oldabs}{\oldabs*}}

\let\oldnorm\norm
\def\norm{\@ifstar{\oldnorm}{\oldnorm*}}
\makeatother

\newcommand{\h}{\mathbf{h}}

\newcommand{\w}{\mathbf{w}}

\newcommand{\z}{\mathbf{z}}

\newcommand{\Z}{\mathbf{Z}}

\newcommand{\Real}{\mbox{$\mathbb{R}$}}
\newcommand{\Compl}{\mbox{$\mathbb{C}$}}

\newcommand{\herm}{\mathrm{H}}

\begin{document}

\title{Empirical Validation of the Impedance-Based RIS Channel Model in an Indoor Scattering Environment}

\author{\IEEEauthorblockN{
Placido Mursia\IEEEauthorrefmark{1},   
Taghrid Mazloum\IEEEauthorrefmark{2},   
Frédéric Munoz\IEEEauthorrefmark{2},    
Vincenzo Sciancalepore\IEEEauthorrefmark{1},      
Gabriele Gradoni\IEEEauthorrefmark{3},\\      
Raffaele D'Errico\IEEEauthorrefmark{2},      
Marco Di Renzo\IEEEauthorrefmark{4},      
Xavier Costa-Pérez\IEEEauthorrefmark{1}\IEEEauthorrefmark{5},     
Antonio Clemente\IEEEauthorrefmark{2},
Geoffroy Lerosey\IEEEauthorrefmark{6}
}                                     
\IEEEauthorblockA{\IEEEauthorrefmark{1}
NEC Laboratories Europe GmbH, Heidelberg, Germany, Email: placido.mursia@neclab.eu}
\IEEEauthorblockA{\IEEEauthorrefmark{2}
CEA-LETI, Univ. Grenoble Alpes,  Grenoble, France}
\IEEEauthorblockA{\IEEEauthorrefmark{3}
University of Surrey, Guildford, United Kingdom}
\IEEEauthorblockA{\IEEEauthorrefmark{4}
Université Paris-Saclay, CNRS, Centrale Supélec, Gif-sur-Yvette, France}  
\IEEEauthorblockA{\IEEEauthorrefmark{5}
i2CAT Foundation and ICREA, Barcelona, Spain}
\IEEEauthorblockA{\IEEEauthorrefmark{6}
Greenerwave, Paris, France}

}

\maketitle

\begin{abstract}
Ensuring the precision of channel modeling plays a pivotal role in the development of wireless communication systems, and this requirement remains a persistent challenge within the realm of networks supported by Reconfigurable Intelligent Surfaces (RIS). Achieving a comprehensive and reliable understanding of channel behavior in RIS-aided networks is an ongoing and complex issue that demands further exploration.
In this paper, we empirically validate a recently-proposed impedance-based RIS channel model that accounts for the mutual coupling at the antenna array and precisely models the presence of scattering objects within the environment as a discrete array of loaded dipoles. To this end, we exploit real-life channel measurements collected in an office environment to demonstrate the validity of such a model and its applicability in a practical scenario. Finally, we provide numerical results demonstrating that designing the RIS configuration based upon such model leads to superior performance as compared to reference schemes.
\end{abstract}

\vskip0.5\baselineskip

\section{Introduction}

Reconfigurable intelligent surfaces (RISs) are widely considered as one of the ongoing revolutions in the network design due to \change{their} property to programmatically alter the propagation properties of the radio environment while keeping the overall manufacturing cost affordable. As such, RISs are a candidate technology for the next-generation wireless networks~\cite{Strinati2021}.
In this regard, one of the main research topics has been on exploring and further developing accurate channel modelling, which is essential to fully unlock the potentials of this technology.

Recently, the authors in~\cite{Gradoni21} have proposed a mutual coupling and unit cell-aware electromagnetically-consistent channel model based on mutual impedances. \change{Therein}, the unit cells of the RIS are modelled as arbitrarily-spaced wire dipoles, whose loads can be tuned to alter the propagation conditions. Such a model has been exploited in~\cite{Qian20} to derive the optimal RIS configuration that maximizes the received power in a single-user and single-antenna setup. Moreover, the authors in~\cite{Abrardo2021} \change{have generalized} this optimization framework for the case of multi-input multiple-output (MIMO) systems. \change{Therein}, the multipath components originated by scattering objects in the environment are modelled \change{as} an additive statistical component. \change{On the other hand, the authors of~\cite{Mursia23}} exploit the discrete dipole approximation (DDA)~\cite{YURKIN2007558} to model the scattering objects present in the environment as loaded wire dipoles, thus effectively capturing its interaction with the transmitted signals. 
Under this setting, the authors \change{of}~\cite{Hassani23} developed a provably convergent and nearly-optimal RIS optimization algorithm based on Gram-Schmidt's orthogonalization method.

In this paper, we go one-step beyond and provide an empirical validation of the model in~\cite{Mursia23} by showcasing how its parameters can be tuned to effectively recreate a real-life environment. To this purpose, we exploit a set of channel measurements~\cite{mazloum2023} performed in an office environment wherein a reflective RIS operating at $28$~GHz~\cite{Popov2021} aids in the communication between a 
transmitter~\cite{7762041} and a single-antenna user equipment (UE), in the presence of several scattering objects. Additionally, we demonstrate how the considered RIS channel model effectively captures such physical propagation scenario with few parameters. Finally, numerical results demonstrate that performing the RIS optimization using such a tailored channel model can bring significant gains in terms of received power at the UE.

\textbf{Notation.} We use bold font lower and upper case for vectors and matrices, respectively. $(\cdot)^\herm$ denotes the hermitian transpose operator, while $j=\sqrt{-1}$ is the imaginary number.


\section{System model}\label{sec:sm}

\begin{figure}[t]
        \center
        \includegraphics[width=0.7\columnwidth]{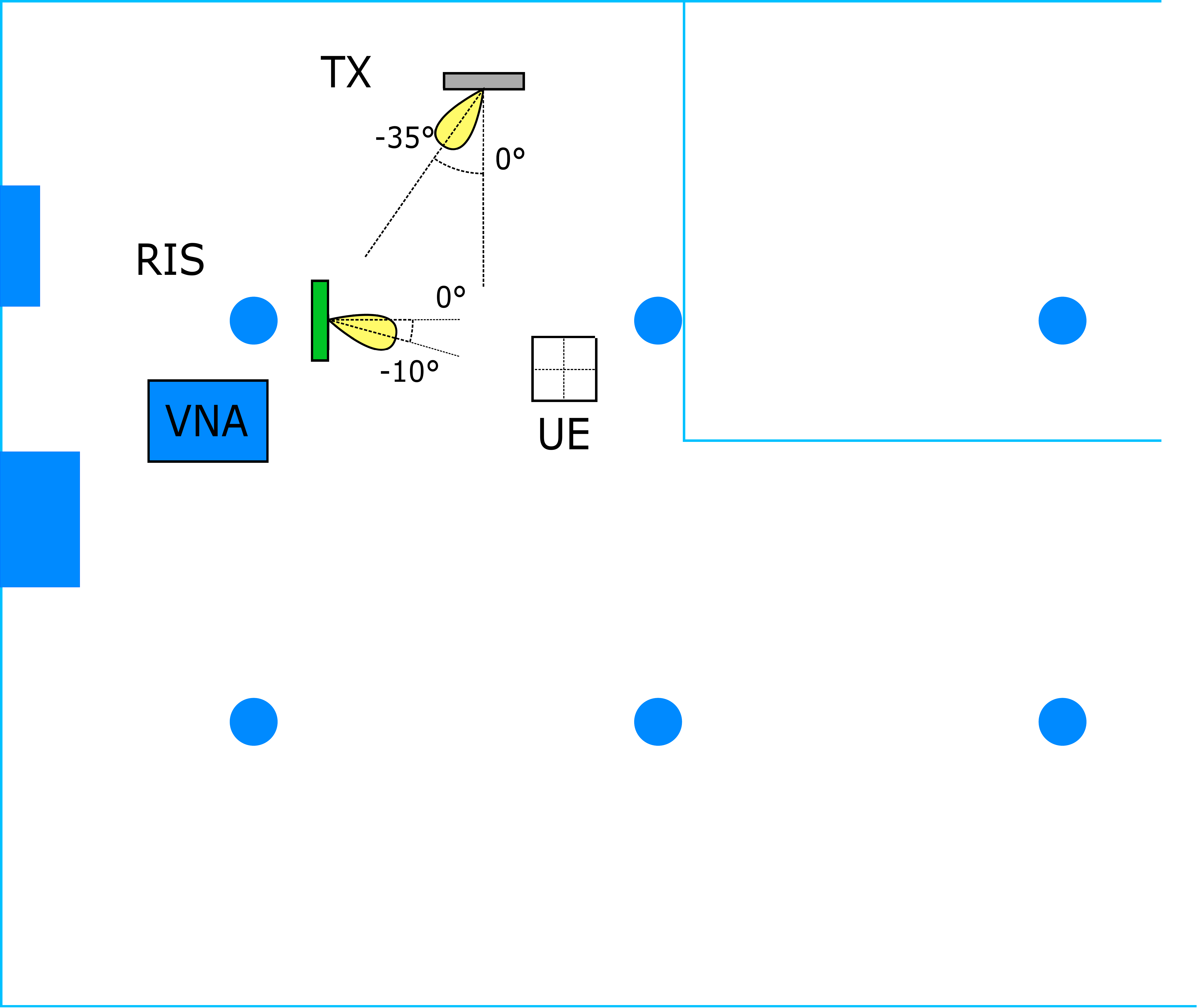}
        \caption{System model}
        \label{fig:Scenario}
\end{figure}

We consider the scenario depicted in Fig.~\ref{fig:Scenario}, wherein a \change{transmitter (TX) equipped with $M=400$ antenna elements is located in the origin of the reference system~\cite{7762041}}, an RIS equipped with $N = 1600$ elements is placed \change{(i.e., its center-point)} at coordinates $\begin{bmatrix}-2 & -3 & 0\end{bmatrix}$~m~\cite{Popov2021}, a single-antenna UE is placed at coordinates $\begin{bmatrix}1 & -3.6 & 0\end{bmatrix}$~m, and a number of scattering objects are present in the environment.
Moreover, we employ the impedance-based RIS channel model in~\cite{Mursia23}, such that the scattering objects in the environment are modelled as discrete arrays of half-wavelength loaded dipoles. Hence, the $1\times M$ channel vector is given by
\begin{align}
    \h^\herm = Z_{\rl}\Big[\z_{\rot}-\z_{\ros}\Big(\Z_{\Ss}+\Z_{\sos}+\Z_{\ris}\Big)^{-1}\Z_{\sot}\Big]\Z_{\tg},\label{eq:h}
\end{align}
where $Z_{\rl}\in\Compl$ ($\Z_{\tg}\in\Compl^{M\times M}$) accounts for the load and self-impedance at the UE (transmitter); $\z_{\rot}\in\Compl^{1\times M}$ includes the mutual impedances between the transmitter and the UE, both directly and indirectly through the scattering objects; $\z_{\ros}\in\Compl^{1\times N}$ ($\Z_{\sot}\in\Compl^{N\times M}$) models the mutual impedance between the RIS and the UE (transmitter), including the effect due to the scattering objects; $\Z_{\Ss}\in \Compl^{N\times N}$ and $\Z_{\sos}\in \Compl^{N\times N}$ represent the self-impedance among the RIS elements and the self and mutual impedance among the scattering objects and the RIS, respectively; lastly, $\Z_{\ris}\in\Compl^{N\times N}$ stands for the diagonal matrix of tunable RIS impedances modelled as $\Z_{\ris} = \mathrm{diag}(R_0+j x_{\ell}), \quad \ell=1,\ldots,N$, where $R_0\geq 0$ is the constant resistance of each RIS load and $x_{\ell}\in \Real$ is the tunable reactance at the $\ell$-th element. 
\change{The precoding vector at the transmitter is denoted by} $\w\in \Compl^{M\times 1}$. As depicted in Fig.~\ref{fig:Scenario}, such vector is defined in order to produce a beam pointing at $-35^\circ$ in azimuth and $0^\circ$ in elevation. Specifically, its beampattern is given in~\cite{7762041}. We remark that the considered measurements collected in~\cite{mazloum2023} involve arrays of patch antennas with a uniform gain of $6$~dBi, while the adopted RIS channel model in~\cite{Mursia23} assumes dipole antennas. Hence, to compensate for the reduced antenna gain, we design the precoder vector such that $\|\w\|^2\leq M \,G$, with $G = 6$~dBi. Moreover, \change{in order to reproduce the setting in~\cite{mazloum2023}}, we fix the RIS configuration in $\Z_{\ris}$ by feeding the channel vector in~\eqref{eq:h} in the absence of both the scattered paths and the direct link, to the \emph{SARIS} algorithm in~\cite{Mursia23} (see Section~\ref{subsubsec:ref_path} for details). The resulting beam at the RIS exhibits a main lobe pointing at $-10^\circ$ in azimuth and $0^\circ$ in elevation. As \change{previously done} for the precoder vector, we then compensate for the patch antenna gain \change{at the RIS} by multiplying the channel coefficient corresponding to the path reflected upon the RIS by $\sqrt{G}$.
Lastly, the receive signal at the UE is thus given by
\begin{align}
    y = \h^\herm \w x + n \in \Compl,
\end{align}
where $x\in\Compl$ is the (known) transmit signal, and $n$ is the additive white Gaussian noise coefficient.

\section{Empirical validation procedure}

In this section, we firstly describe the considered measurements collected in the real-life environment described in Fig.~\ref{fig:Scenario}. Subsequently, focusing on the channel model in~\eqref{eq:h}, we describe the procedure to fit the associated relevant parameters in order to match the underlying physical propagation environment. 

\subsection{Channel measurements}\label{sec:measurements}

In the indoor environment depicted in Fig.~\ref{fig:Scenario}, the propagation channel was measured using a vector network analyzer (VNA) operating within the frequency range of $25$-$35$~GHz.  The setup included the use of an RIS as an extender, a 
TX, and a wideband monopole antenna serving as UE.  All of these terminals were placed at a height of about $1.6$~m above the ground, corresponding to the azimuth plane. Further details on the measurement setup and the analysis of the channel characteristics are found in~\cite{mazloum2023}. The TX was configured to steer the beam towards the RIS, which, in turn, was configured to redirect the received beam towards the intended UE. The UE, placed on a 2D positioner, \change{was moved on} a $3 \times 3$ spatial grid \change{of positions, with a step equivalent to half a wavelength, thus emulating a $3 \times 3$ antenna array}. Across a bandwidth of $2$~GHz, this spatial grid allows the extraction of the multipath components (MPCs) in both the time delay (or alternatively the distance) and the angular domains, where the latter corresponds specifically to the azimuth angle-of-arrival (AoA) at the UE-side. This is achieved by applying a wideband high resolution algorithm such as the Space-Alternating Generalized Expectation-Maximization algorithm (SAGE)~\cite{UWB_SAGE}.

\change{The TX and the RIS} operate within the Ka band at the resonance frequency of $28$~GHz.  The linearly-polarized TX \cite{7762041} \change{consists} of $20 \times 20$ unit cells (UCs) \change{and is} illuminated with a $10$-dBi horn antenna. The employed RIS \cite{Popov2021} is an assembly of four rectangular lattices of $20 \times 20$ UCs each, resulting on a total of $40 \times 40$ UCs. The \change{one-bit} UCs \change{of the TX and the RIS} are arranged at regular intervals of half a wavelength \change{and the configuration is realized} through electronic control of integrated PIN diodes. This enables ON and OFF states with a relative phase-shift \change{of approximately $180^{\circ}$ at the chosen operating frequency}. Further details on the $1$-bit UC architecture and performance are provided in~\cite{7762041} for the TX and in~\cite{Popov2021} for the RIS.

\begin{figure}[t]
        \center
        \includegraphics[width=0.8\columnwidth]{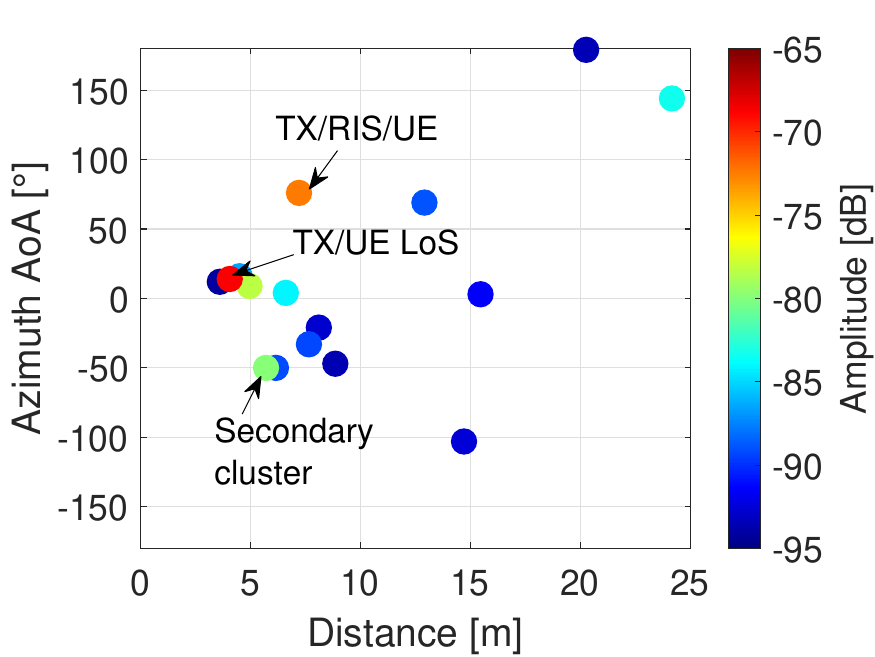}
        \caption{Heatmap of the time delay (or alternatively the distance) and azimuth AoA for the multipath in the considered scenario as perceived by the UE.}
        \label{fig:Heatmap}
\end{figure}

The heatmap of the MPCs as perceived by the UE, extracted using the SAGE algorithm, is illustrated in Fig.~\ref{fig:Heatmap}, where each path is graphically represented by a point on the angle-distance axes, and its color is related to its amplitude (in dB).  The identification of the MPCs on the heatmap is simplified with geometrical calculations based on the dimension of the floor plan. Accordingly, the MPC \change{reflected} by the RIS appears at $7.2$~m with an AoA of $76^\circ$, and the direct TX to UE path is located at $4$~m with an AoA of $14^\circ$. Their powers are respectively $-72.4$~dB and $-68.7$~dB. \change{Also}, the \change{strongest} scattered path \change{due to} the environmental objects is represented by the one corresponding to a distance of $5.69$~m, AoA of $-50^\circ$ and a power of $-79.8$~dB. All other MPCs are due to scatterers that are far-away from the UE or to measurement noise. In this paper for model validation we limit our analysis to the main considered MPCs summarized in Table~\ref{tab:MPCs}.

\begin{table}[t!]
\caption{Main MPCs in terms of distance travelled, AoA at the UE and measured power.}
\label{tab:MPCs}
\centering
\resizebox{1\linewidth}{!}{%
\renewcommand{\arraystretch}{1.0}
\begin{tabular}{|c|c|c|c|}
\hline
\cellcolor[HTML]{EFEFEF} \textbf{MPC} & \textbf{Distance} &\textbf{AoA} & \textbf{Power}\\
\hline 
 \cellcolor[HTML]{EFEFEF} Direct path (TX/UE LoS) & $4$~m & $14^\circ$ & $-68.7$~dB \\
\hline
 \cellcolor[HTML]{EFEFEF} Reflected path (TX/RIS/UE) & $7.2$~m & $76^\circ$ &  $-72.4$~dB\\
\hline 
\cellcolor[HTML]{EFEFEF} Secondary Cluster & $5.69$~m  &
 $-50^\circ$ & $-79.8$~dB \\
\hline
\end{tabular}%
}
\renewcommand{\arraystretch}{1}
\end{table}

\subsection{Channel model fitting}\label{sec:model_fit}

In order to tune the parameters in the model described in Section~\ref{sec:sm} \change{and replicate the setup in~\cite{mazloum2023}}, we distinguish three cases, namely \textit{i)} the direct path from the transmitter to the receiver  i.e. the TX/UE Line of Sight (LoS) path, \textit{ii)} the reflected path from the transmitter to the UE via the \change{RIS}, i.e. the TX/RIS/UE path,   and \textit{iii)} the scattered path from the transmitter to the receiver via a cluster of reflectors in the vicinity of the UE.

\subsubsection{Direct path}\label{subsubsec:dir_path}
Regarding the direct path from the transmitter to the UE, we modify the channel vector as
\begin{align}
    \h_{d}^\herm = Z_{\rl}\z_{\RT}\Z_{\tg},
\end{align}
where $\z_{\RT}$ is the mutual impedance between the transmitter and the UE. As stated above, the transmitter configuration is set as described in~\cite{7762041}.
\subsubsection{Reflected path}\label{subsubsec:ref_path}
In this case, we consider only the path that reaches the UE from the transmitter via the RIS. Hence, we modify the channel vector in~\eqref{eq:h} as
\begin{align}
    \h_{\ris}^\herm = -Z_{\rl}\Big[\z_{\rs}\Big(\Z_{\Ss}+\Z_{\ris}\Big)^{-1}\Z_{\ST}\Big]\Z_{\tg},\label{eq:h_ris}
\end{align}
with $\z_{\rs}$ and $\Z_{\ST}$ the mutual impedance between the receiver and the RIS and between the latter and the transmitter in the absence of scattering objects, respectively. Here, all parameters can be computed given the (fixed) location of the transmitter, the RIS and the receiver~\cite{Gradoni21}. The TX configurations is set as described in Section~\ref{sec:sm}, while the RIS configuration is obtained by feeding~\eqref{eq:h_ris} to the SARIS algorithm~\cite{Mursia23}.

\subsubsection{Secondary cluster}\label{subsubsec:scat_path}
Lastly, we focus on the path that reaches the UE from the \change{TX} via a cluster of scatterers, which is located in the vicinity of the UE. In this case, the channel vector is given by
\change{
\begin{align}
    \h_{ref}^\herm = -Z_{\rl}\z_{\ro}(\Z_{\oo}+\Z_{\us})^{-1}\z_{\ot}\Z_{\tg}.
\end{align}
}
As described in~\cite{Mursia23}, \change{$\z_{\ro}$ represents the mutual impedance between the UE and the cluster of scatterers, $\Z_{\oo}$ the self and mutual impedances among the scatterers, $\Z_{\us}$ the diagonal matrix modelling the load attached to each scattering dipole, and $\z_{\ot}$ stands for the mutual impedance between the cluster of scatterers and the TX. Note that the dimension of such quantities depends on $N_c$, i.e., the number of scattering dipoles, which needs to be fitted to accturately represent the cluster}. In particular, based upon the floor plan of the channel measurements described in Section~\ref{sec:measurements}, such \change{cluster} consists of a cylindrical pillar with a width of approximately $1$~m and a height of approximately $3$~m. However, given the narrow shape in 3D of the beam pattern at the transmitter, it is sufficient to model only a small section of $16\times 8$~cm. Specifically, by applying the DDA~\cite{YURKIN2007558}, we model such object as a cylindrical array of radius $0.5$~m, consisting of \change{$N_c = N_x\times N_y$} loaded dipoles of length $h = \lambda/4$ whose inter-element distance is given by $d$, with $N_x$ the number of dipoles on the horizontal plane, $N_y$ the number of dipoles on the vertical plane, and $\lambda$ the signal wavelength. The load on each $t$-th dipole is given by \change{$Z_{\us t} = y_0 + j x_t$}, where $y_0\in\Real$ is constant over the array, and it depends on the specific absorption and reflection properties of the material, while $x_t\in\Real$ models the scattering properties of the object. To this end, the values of $N_x$, $N_y$, $d$, $\{x_t\}_{t=1}^{N_x\times N_y}$, as well as the location of the object need to be fitted to reproduce the results in Table~\ref{tab:MPCs}.

\section{Numerical results and discussion}

In order to validate the effectiveness of the channel model in~\eqref{eq:h} and~\cite{Mursia23}, we evaluate the (normalized) power received at the UE-side in dB, which is given by
\begin{align}
    P = 10\log_{10}(|\h^\herm \w|^2),
\end{align}
in the various cases highlighted in Section~\ref{sec:model_fit}. The obtained results for the reflected and direct paths are shown in Table~\ref{tab:Simul}, demonstrating an error of just $1.16\%$ and $0.35\%$ for the two cases, respectively. 
For the case of the path \change{including} the cylindrical reflector, we generate $10^3$ different combinations of the reactance $x_t$ for the load of each $t$-th dipole in the cylindrical array by drawing values from the uniform distribution $\mathcal{U}[-330,100]$. Furthermore, we vary the total number of dipoles $N_x\times N_y$ with fixed ratio $N_y =2 N_x$, the inter-element distance $d$, and the real part of the dipole loads $y_0$. For each combination, we select the maximum value obtained over the random reactance. 

\begin{figure}[t]
        \center
        \includegraphics[width=0.8\columnwidth]{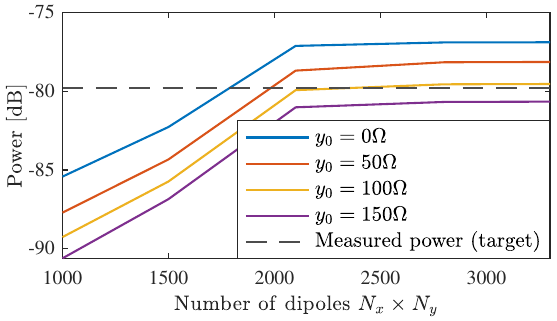}
        \caption{Received power versus the number of dipoles in the cylindrical array with inter-element spacing of $d=\lambda/4$, and for different values of $y_0$.}
        \label{fig:Close_Cluster}
\end{figure}

In Fig.~\ref{fig:Close_Cluster}, we show the receive power at the UE versus the total number of dipoles for the case of $d=\lambda/4$, and for different values of $y_0$. We thus conclude that \change{approximately} $2450$ dipoles, corresponding to a $35\times 70$ configuration, with $y_0 = 100~\Omega$ is a precise enough fitting of the measured data, resulting in an error of $0.14\%$. Moreover, in Table~\ref{tab:P_vs_d}, we provide the simulated received power for the case of a $35\times 70$ cylindrical array with $y_0=100 ~\Omega$ for different values of the inter-element $d$, thus confirming that a dense array accurately represents the physical propagation properties of a real-life object.

\change{While so far we have successfully emulated the measurements in~\cite{mazloum2023} by configuring the RIS as described in Section~\ref{sec:sm}, in the following we demonstrate how the algorithm in~\cite{Mursia23} can lead to superior performance by designing an RIS configuration that explicitly interacts with the scattering environment and creates constructive interference at the UE. In this regard,} we set all channel parameters according to the values that provide the best fitting of the physical propagation environment, as described above, and compare the RIS configuration used during the measurements described in Section~\ref{sec:measurements} versus the SARIS algorithm in~\cite{Mursia23}. As shown in Table~\ref{tab:SARIS}, the proposed RIS optimization algorithm achieves significant gains in the received power thanks to an accurate modelling and exploitation of the physical properties of the channel.

\begin{table}[t!]
\caption{Simulated MPCs for the reflected and direct paths versus the measured data.}
\label{tab:Simul}
\centering
\resizebox{1\linewidth}{!}{%
\renewcommand{\arraystretch}{1.0}
\begin{tabular}{|c|c|c|}
\hline
\cellcolor[HTML]{EFEFEF} \textbf{MPC} & \textbf{Measured power} & \textbf{Simulated power}\\
\hline
 \cellcolor[HTML]{EFEFEF} Direct path & $-68.7$~dB & $-68.5$~dB \\
\hline 
 \cellcolor[HTML]{EFEFEF}  Reflected path  &  $-72.4$~dB & $-73.3$~dB\\
\hline
\end{tabular}%
}
\renewcommand{\arraystretch}{1}
\end{table}

\begin{table}[t!]
\caption{Simulated MPC involving the close-by reflector versus different inter-element spacings $d$ in a $35\times 70$ cylindrical array and with $y_0 = 100~\Omega$.}
\label{tab:P_vs_d}
\centering
\resizebox{1\linewidth}{!}{%
\renewcommand{\arraystretch}{1.0}
\begin{tabular}{|c|c|c|c|}
\hline
\cellcolor[HTML]{EFEFEF} \textbf{Simulated power} & \textbf{Inter-element spacing $d$} & \cellcolor[HTML]{EFEFEF} \textbf{Simulated power} & \textbf{Inter-element spacing $d$}\\
\hline
 \cellcolor[HTML]{EFEFEF}  $-113.85$~dB & $5\lambda$ & \cellcolor[HTML]{EFEFEF} $-89.45$~dB & $\lambda$ \\
\hline 
 \cellcolor[HTML]{EFEFEF} $-86.64$~dB & $0.5\lambda$ &  \cellcolor[HTML]{EFEFEF} $-79.93$~dB & $0.25\lambda$ \\
\hline
\end{tabular}%
}
\renewcommand{\arraystretch}{1}
\end{table}

\begin{table}[t!]
\caption{Received power in the considered scenario under different RIS configurations.}
\label{tab:SARIS}
\centering
\resizebox{1\linewidth}{!}{%
\renewcommand{\arraystretch}{1.0}
\begin{tabular}{|c|c|}
\hline
\cellcolor[HTML]{EFEFEF} \textbf{RIS configuration} & \textbf{Simulated power}\\
\hline
 \cellcolor[HTML]{EFEFEF} $-10^\circ$ azimuth, $0^\circ$ elevation &  $-63.27$~dB \\
\hline 
 \cellcolor[HTML]{EFEFEF} SARIS~\cite{Mursia23} & $-52.34$~dB \\
\hline
\end{tabular}%
}
\renewcommand{\arraystretch}{1}
\end{table}

\section{Conclusion}

In this paper, we shed \change{experimentally validate} a recently-introduced electromagnetic-compliant channel model for RIS, which takes into account mutual coupling effects and considers the presence of scattering objects.
To this end, we have validated its \change{accuracy in} representing the physical characteristics of real-world scenarios. Interestingly, we have fine-tuned the \change{parameters of model} using a dataset of channel measurements collected from an office environment. \change{By recreating} the channel propagation \change{with the DDA, we have demonstrated that the proposed} modeling can be leveraged to derive enhanced RIS configurations, leading to substantial improvements in the overall system performance.


\section*{Acknowledgment}

This work was supported in part by the EU H2020 RISE-6G project under grant 101017011.

\bibliographystyle{IEEEtran}
\bibliography{refs}

\begin{thebibliography}{10}
\providecommand{\url}[1]{#1}
\csname url@samestyle\endcsname
\providecommand{\newblock}{\relax}
\providecommand{\bibinfo}[2]{#2}
\providecommand{\BIBentrySTDinterwordspacing}{\spaceskip=0pt\relax}
\providecommand{\BIBentryALTinterwordstretchfactor}{4}
\providecommand{\BIBentryALTinterwordspacing}{\spaceskip=\fontdimen2\font plus
\BIBentryALTinterwordstretchfactor\fontdimen3\font minus \fontdimen4\font\relax}
\providecommand{\BIBforeignlanguage}[2]{{%
\expandafter\ifx\csname l@#1\endcsname\relax
\typeout{** WARNING: IEEEtran.bst: No hyphenation pattern has been}%
\typeout{** loaded for the language `#1'. Using the pattern for}%
\typeout{** the default language instead.}%
\else
\language=\csname l@#1\endcsname
\fi
#2}}
\providecommand{\BIBdecl}{\relax}
\BIBdecl

\bibitem{Strinati2021}
E.~Calvanese~Strinati, G.~C. Alexandropoulos, V.~Sciancalepore, M.~Di~Renzo, H.~Wymeersch, D.-T. Phan-huy, M.~Crozzoli, R.~D'Errico, E.~De~Carvalho, P.~Popovski, P.~Di~Lorenzo, L.~Bastianelli, M.~Belouar, J.~E. Mascolo, G.~Gradoni, S.~Phang, G.~Lerosey, and B.~Denis, ``Wireless environment as a service enabled by reconfigurable intelligent surfaces: The {RISE-6G} perspective,'' in \emph{IEEE EuCNC/6G Summit}, 2021, pp. 562--567.

\bibitem{Gradoni21}
G.~Gradoni and M.~{Di Renzo}, ``{End-to-End Mutual Coupling Aware Communication Model for Reconfigurable Intelligent Surfaces: An Electromagnetic-Compliant Approach Based on Mutual Impedances},'' \emph{IEEE Wireless Communications Letters}, vol. 2337, no.~c, pp. 1--5, 2021.

\bibitem{Qian20}
X.~Qian and M.~{Di Renzo}, ``Mutual coupling and unit cell aware optimization for reconfigurable intelligent surfaces,'' \emph{IEEE Wireless Commun. Lett.}, vol.~10, no.~6, pp. 1183--1187, 2021.

\bibitem{Abrardo2021}
A.~Abrardo \emph{et~al.}, ``{MIMO} interference channels assisted by reconfigurable intelligent surfaces: Mutual coupling aware sum-rate optimization based on a mutual impedance channel model,'' \emph{IEEE Wireless Commun. Lett.}, vol.~10, no.~12, pp. 2624--2628, 2021.

\bibitem{Mursia23}
P.~Mursia, S.~Phang, V.~Sciancalepore, G.~Gradoni, and M.~Di~Renzo, ``Saris: Scattering aware reconfigurable intelligent surface model and optimization for complex propagation channels,'' \emph{IEEE Wireless Communications Letters}, pp. 1--1, 2023.

\bibitem{YURKIN2007558}
M.~A. Yurkin and A.~G. Hoekstra, ``The discrete dipole approximation: {A}n overview and recent developments,'' \emph{J. Quantitative Spectroscopy and Radiative Transfer}, vol. 106, no.~1, pp. 558--589, 2007.

\bibitem{Hassani23}
\BIBentryALTinterwordspacing
H.~E. Hassani, X.~Qian, S.~Jeong, N.~S. Perovi{\'{c}}, M.~{Di Renzo}, P.~Mursia, V.~Sciancalepore, and X.~Costa-P{\'{e}}rez, ``{Optimization of RIS-Aided MIMO -- A Mutually Coupled Loaded Wire Dipole Model},'' 2023. [Online]. Available: \url{http://arxiv.org/abs/2306.09480}
\BIBentrySTDinterwordspacing

\bibitem{mazloum2023}
T.~Mazloum, L.~Santamaria, F.~Munoz, A.~Clemente, J.-B. Gros, Y.~Nasser, M.~Odit, G.~Lerosey, and R.~D'Errico, ``Impact of multiple ris on channel characteristics: An experimental validation in ka band,'' in \emph{2023 Joint European Conference on Networks and Communications \& 6G Summit (EuCNC/6G Summit)}, 2023, pp. 13--18.

\bibitem{Popov2021}
V.~Popov, M.~Odit, J.-B. Gros, V.~Lenets, A.~Kumagai, M.~Fink, K.~Enomoto, and G.~Lerosey, ``Experimental demonstration of a {mmWave} passive access point extender based on a binary reconfigurable intelligent surface,'' \emph{Frontiers in Communications and Networks}, vol.~2, 2021.

\bibitem{7762041}
L.~Di~Palma, A.~Clemente, L.~Dussopt, R.~Sauleau, P.~Potier, and P.~Pouliguen, ``Circularly-polarized reconfigurable transmitarray in ka-band with beam scanning and polarization switching capabilities,'' \emph{IEEE Transactions on Antennas and Propagation}, vol.~65, no.~2, pp. 529--540, 2017.

\bibitem{UWB_SAGE}
K.~Haneda and J.-I. Takada, ``An application of sage algorithm for uwb propagation channel estimation,'' in \emph{IEEE Conference on Ultra Wideband Systems and Technologies, 2003}, 2003, pp. 483--487.

\end{thebibliography}

\end{document}